\documentclass{aa} 
\usepackage[varg]{txfonts}
\usepackage{graphicx}
\pdfoutput=1

\begin{document}

\title{Time variability in the bipolar scattered light nebula of L1527 IRS: A possible warped inner disk}

\titlerunning{Time variability of L1527 IRS}

\author{Brian T. Cook$^1$
\and John J. Tobin$^2$
\and Michael F. Skrutskie$^3$
\and Matthew J. Nelson$^3$}

\authorrunning{Cook, Tobin, Skrutskie, and Nelson}

\institute{Leiden Observatory \\
         2333 CA Leiden, 
         Netherlands
\and National Radio Astronomy Observatory \\
         Charlottesville, VA, 22903, USA
\and Astronomy Department, University of Virginia \\ 
     Charlottesville, VA 22901, USA}

\date{Received March 6, 2019 /
Accepted April 29, 2019} 

\abstract {The bipolar outflows associated with low-mass protostars create cavities in the infalling envelope. These cavities are illuminated by the central protostar and inner disk, creating a bipolar scattered light nebula at near-infrared and mid-infrared wavelengths. The variability of the scattered light nebula in both total intensity and intensity as a function of position in the scattered light nebula can provide important insights into the structure of the inner disk that cannot be spatially resolved.} {We aim to determine the likelihood that a warped inner disk is the origin of the surface brightness variability in the bipolar scattered light nebula associated with L1527 IRS.} {We present results from near-IR imaging conducted over the course of seven years, with periods of monthly cadence monitoring. We used Monte Carlo radiative
transfer models to interpret the observations.} {We find a time varying, asymmetrical brightness in the scattered light nebulae within the outflow cavities of the protostar. Starting in 2007, the surface brightnesses of the eastern and western outflow cavities were roughly symmetric. Then, in 2009, the surface brightnesses of the cavities were found to be asymmetric, with a substantial increase in surface brightness and a larger increase in the eastern outflow cavity. More regular monitoring was conducted from 2011 to 2014, revealing
a rotating pattern of surface brightness variability in addition to a slow change of the eastern and western outflow cavities toward symmetry, but still not as symmetric as observed in 2007. We find that an inner disk warp is a feasible mechanism to produce the rotating 
        pattern of surface brightness variability.} {}


\maketitle

\section{Introduction}

The formation of stars occurs in dense cores generally within molecular clouds. It is within these
regions of high density
where gravity overwhelms supporting forces (e.g., turbulence, magnetic fields, and 
thermal pressure) and the core begins to collapse. A disk is formed 
around a protostar during collapse due to the conservation of angular 
momentum and the protostar and disk work together to drive a bipolar outflow, probably
through magneto-centrifugal acceleration \citep[e.g.,][]{blandford1982}. 
The bipolar outflow then carves out conical cavities in the envelope that can
be probed with observations of CO molecular gas \citep[e.g.,][]{snell1980,arce2007}, but also
at near-infrared wavelengths where light from the protostar and inner disk 
escape through the outflow cavities and scatter on dust grains in the outflow
cavity and along the cavity walls. 

The system L1527 IRS is located within the Taurus molecular cloud at a distance of $\sim$140~pc \citep{torres2007}. It is a Class 0/I system that is characterized by its dense, presumably infalling,
envelope of gas and dust \citep{andre1993}. Its exact classification is somewhat uncertain
given that it is viewed edge-on, which could make the system look younger \citep{tobin2008}.
Regardless of its exact classification, it has a well-ordered
structure that has made it a prototype for studies of protostellar collapse \citep{ohashi1997,sakai2014}
and protostellar disk formation \citep{tobin2012,ohashi2014}. Within such a system,
radiation from the protostar and disk is absorbed by the dusty envelope and re-emitted 
at longer wavelengths \citep{wood2001}, in addition to scattering at shorter wavelengths. The
shorter wavelength scattering gives rise to scattered light nebulae that can be prominent
toward protostars at near to mid-infrared wavelengths \citep[e.g.,][]{seale2008}.

Scattered light nebulae have long been used to discern the behavior of the emitting object; an early example is Hubble's work on NGC 2261 \citep[also known as R Mon or Hubble's Variable Nebula, ][]{hubble1917}. 
These scattered light images can complement millimeter studies of more optically 
thin dust emission in order to probe the envelope and disk structure of protostars and 
pre-main sequence stars \citep{guilloteau2008,wolf2008,tobin2012,segura-cox2016,sheehan2017}; scattered  light images on envelope scales can also show effects due to changes of the  illumination source, the inner disk, and protostar at the heart of the system \citep{tobin2008,connelley2009}.

The inner disk structure can affect the illumination 
and brightness of the outflow cavities because the inner disk rim emits
the majority of photons at wavelengths that can be detected in scattered light for protostars \citep{whitney20031, muzerolle2003}. A natural cause of variability in the
scattered light nebulae could be a rotating warped disk, and astrophysical disks are often warped. This occurs when an initially planar disk is misaligned to a 
component of the potential (e.g., the orbit of a companion star), or when 
the disk becomes unstable to tilting by tidal \citep{lubow1992} or 
radiation effects \citep{pringle1996, nixon2016}. In \citet{watson2007}, 
the asymmetry of HH 30's circumstellar disk is explored. They found that 
the brightness of the upper nebula could vary by a factor of 0.2 mag 
in just two days. While analyzing the habitable zone of T Tauri stars in the 
mid-infrared, \citet{ke2012} postulate that a magnetized and 
misaligned host star can cause a periodic warp that in turn leads to a 
periodic modulation of the spectral energy distribution. The provided 
explanation for a warp of this kind is either a companion star or a dynamic 
interface between the star's magnetic field and the disk, resulting in a 
variable scale height. 

The variability of L1527 IRS was first recognized 
in multi-epoch images from the \textit{Spitzer Space Telescope} \citep{tobin2008} and it
has been further characterized in ground-based monitoring conducted over the 
course of several years. Section 2 
outlines the observations of L1527 IRS from 2007 to 2014, 
while Sect. 3 is devoted to discussing the observed characteristics of 
the outflow cavities. We explore the possible origin of the scattered
light variability by introducing changes to the inner disk structure 
in Sect. 4 in an effort to qualitatively match the observations and provide
insights into the small-scale structure of the protostellar disk. 
We discuss our results in Sect. 5 and present our conclusions in Sect. 6.

\section{L1527 observations}

The observations were conducted with the FanCam near-infrared camera on 
the 31 inch Tinsley reflector operated by the University of Virginia 
and located at Fan mountain near Charlottesville, Virginia. The camera 
features an 8.7' $\times$ 8.7' field of view on a 1024 $\times$ 1024 Teledyne 
Imaging Sensors HAWAII-1 detector array, and was designed for use from $\sim$1 to 2.5 $\mu$m 
\citep{kanneganti2009}. A small telescope with infrared instrumentation is 
capable of observations over several years, which is not as easily 
achieved at larger observatories. Thus, the FanCam camera was ideal for 
these observations. L1527 IRS was observed on multiple nights from Fall 
2011 to Spring 2014 (see Table \ref{table:observation-log}). We also include previous imaging from the TIFKAM instrument on the 2.4m Hiltner telescope at the MDM\footnote{Originally named the Michigan-Dartmouth-MIT observatory, the acronymn is now the standard name as various institutions have joined or left the consortium.} observatory
in 2007 \citep{tobin2008}, and an additional observation from MDM in December 2009 is also
presented here.

The data were reduced using standard methods for ground-based near-infrared imaging, using
the \textit{upsqiid} package of the Image Reduction and Analysis Facility (IRAF). 
We used twilight sky flat fields or skyflats constructed from the observations. 
The FanCam observations were conducted in a ten-point dither pattern with L1527 IRS centered 
on one quadrant of the detector for each dither pattern, rotating through the quadrants
after each complete dither pattern. The centering of L1527 IRS in 
different quadrants for each dither pattern enabled the preceding and 
following dither patterns to be median combined to produce a sky intensity image
that was then subtracted from the data. Each integration was 30 seconds and the flat-fielded,
sky-subtracted images were median combined to produce a single mosaic. 
Often the final mosaic image from a single night of observations consists of 100 to 200 individual frames. The 
frames were aligned by centroiding on a star that was present in all observations. \citet{tobin2008}
describe the data reduction process with \textit{upsqiid} in more detail. We observed primarily
in the Ks-band filter because the scattered light nebula of L1527 is best detectable 
at this wavelength in ground-based observations.

\section{Results}

\begin{figure*}
        \centering
        \includegraphics[width=1.0\linewidth]{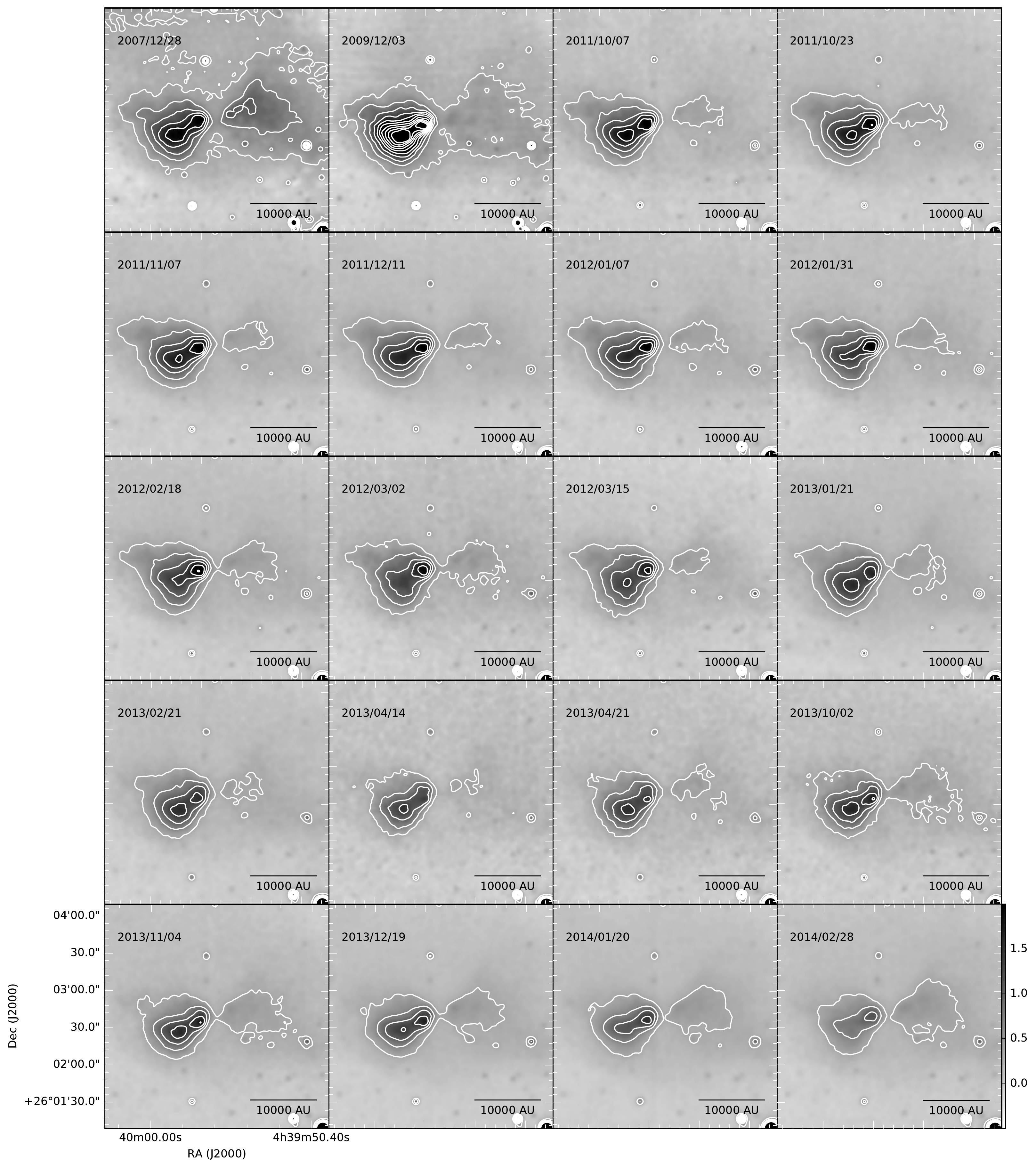}
    \caption{Observations of L1527 in Ks-band from December 2007 to April 2014 shown as inverse
    gray scale. 
The contour levels are $3.5 \times n$ where $n$ is the set of numbers 0.1, 0.2, $\dots$, 1.0. As stated in Sect. 3, the primary features 
of the scattered light nebulae during this observation run are the western
lobe's emergence and the gradual transition toward a more symmetric brightness 
distribution of the cavities. The color bar shown in the lower right panel is in 
units of MJy~sr$^{-1}$.}
        \label{fig:ObservationGrid}
\end{figure*}

\begin{figure*}[h]
        \centering
        \includegraphics[width=0.8\linewidth]{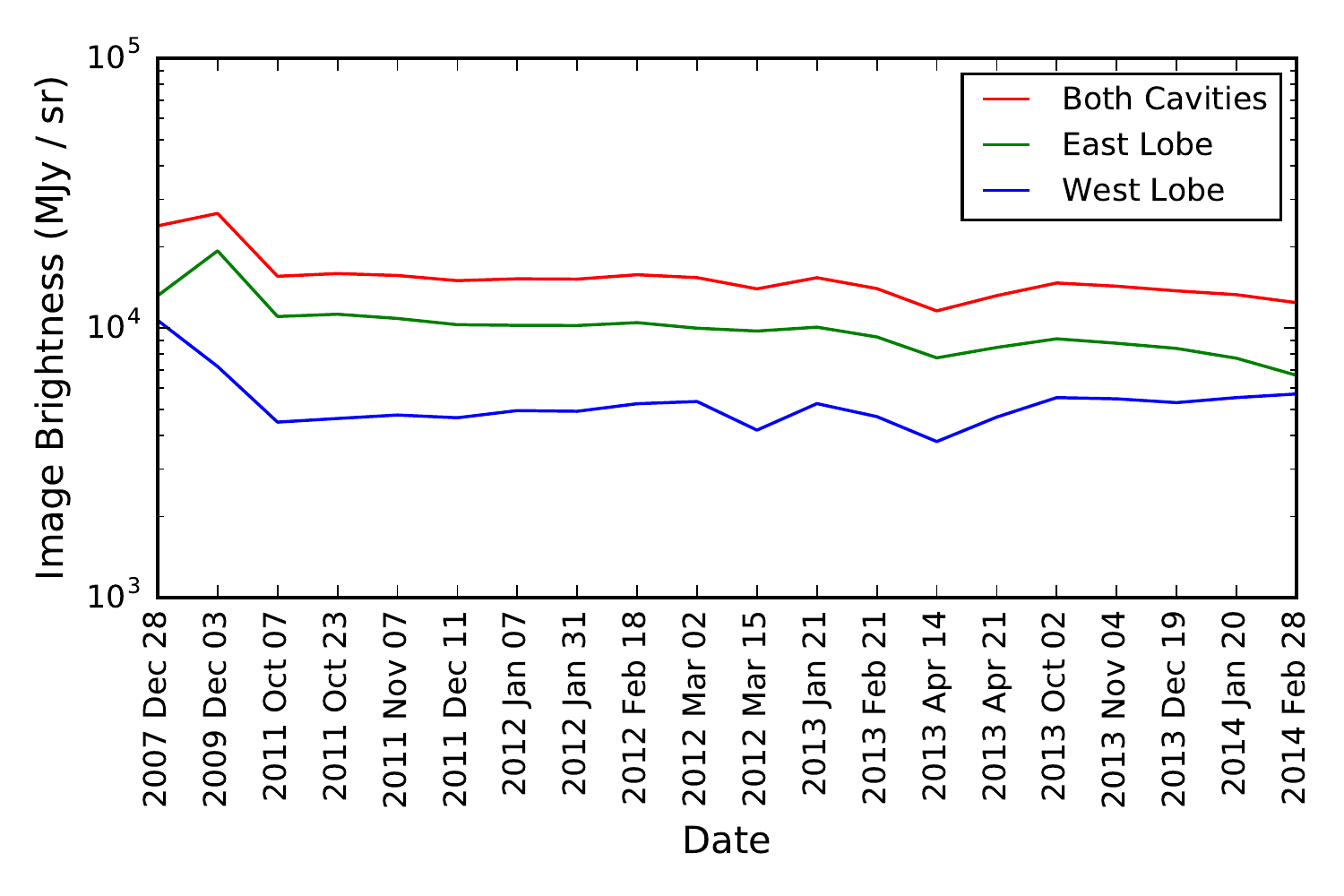}
    \caption{Plot of image brightness as a function of time, indicating that the eastern lobe is indeed brighter for the entire observational run.}
        \label{fig:BrightnessPlot}
\end{figure*}

The near-infrared images toward L1527 IRS are shown in Fig. \ref{fig:ObservationGrid}. The 
outflow cavities of L1527 IRS are oriented in almost exactly the east-west direction. The system is also
oriented with the equatorial plane of the disk viewed nearly edge-on \citep{tobin2008,tobin2010}, making the 
outflow cavities on both sides of the protostar visible through comparable amounts of
extinction, but the eastern side of the outflow cavity remains brighter throughout our observations.
Only in the 2007 observation and final observation in 2014 do they appear at nearly equal brightness, as shown in Fig. \ref{fig:BrightnessPlot}.
It is still not completely clear which side of the outflow cavity is tilted toward our line of sight. 
For a spherical envelope, one would assume that the brighter outflow cavity would be tilted toward us,
and the upper layers of the disk from \citet{tobin2010} show the eastern disk surface to appear brighter
at 3.8 $\mu$m. There is also substantial
overlap between red and blue-shifted CO emission on either side of the outflow cavity \citep{bontemps1996}.
However, by modeling the envelope kinematics \citet{oya2015} suggest that the western outflow cavity 
is tilted slightly toward us, but only by about 5\degr.


The first two observations from 2007 and 2009 in Fig. \ref{fig:ObservationGrid} are substantially different,
which, along with the variability detected with \textit{Spitzer} in \citet{tobin2008},
establishes the variability of the protostellar system and feasibility of characterizing it
with ground-based observations. However, as these observations took place 
several years before the rest (and two years apart), it is difficult to place them within the context of 
any periodic phenomena of L1527 IRS. The surface brightness of the eastern lobe in 2009 is 
significantly greater than at later times. When monthly observations began 
in October 2011, the overall intensity decreased to $\sim$70\% of the 2007 brightness 
and steadily decreased from 2011 to 2014. 

The evolution of the western lobe is more peculiar than that of the eastern lobe. 
In 2009 and 2011, it is only marginally detected and remains faint through 2012 and early 2013.
The brightness of the western lobe began to increase in April 2013 and remained in a brighter state
through the end of observations in early 2014, but it never had a higher surface brightness than the eastern lobe.

\begin{figure*}
        \centering
        \includegraphics[width=1.0\linewidth]{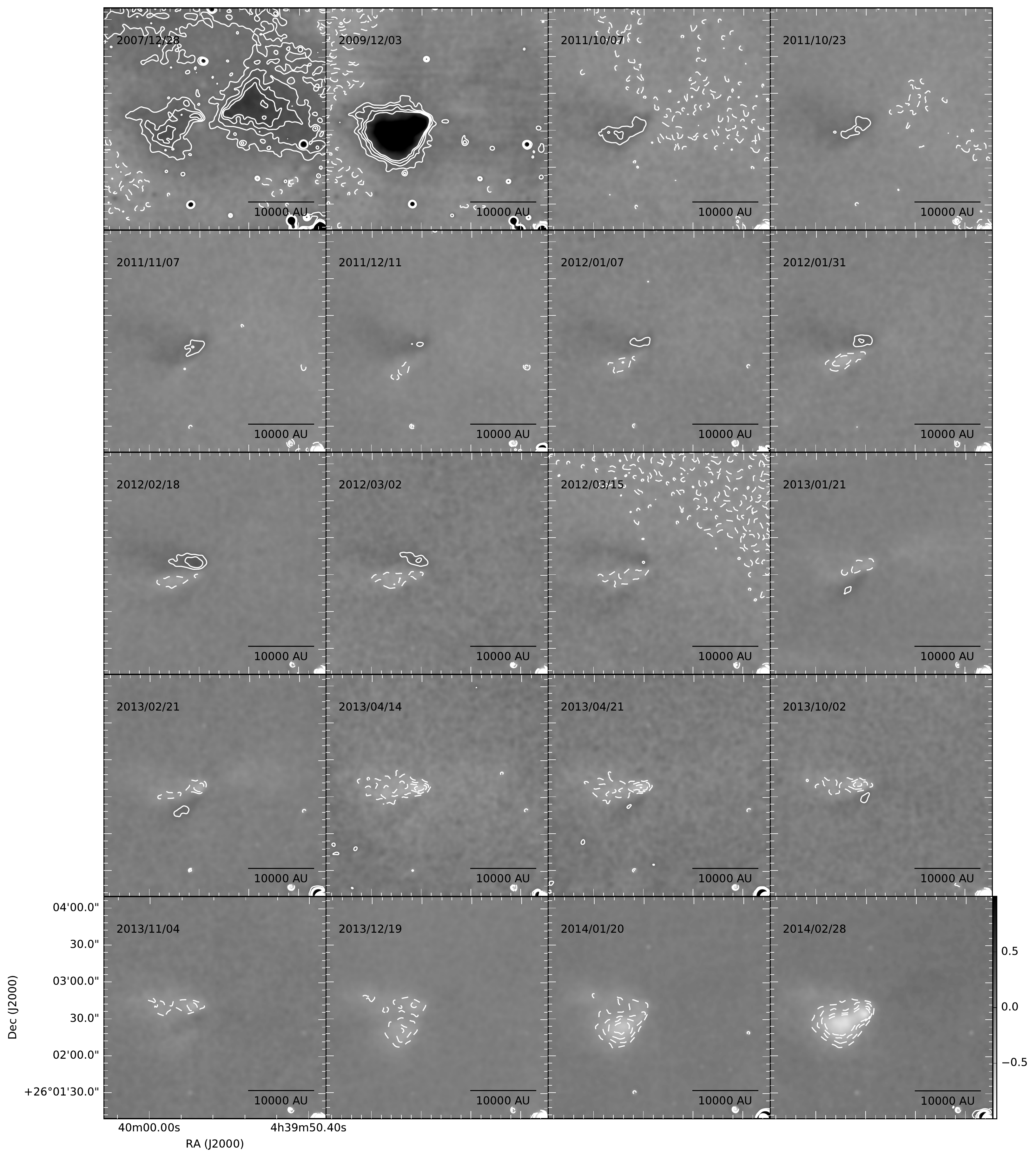}
    \caption{Observations with the average image of L1527 subtracted out, December 2007 to April 2014 displayed as inverse grayscale. 
The contour levels are -0.45, -0.35, -0.25, -0.15, 0.15, 0.25, 0.35, 0.45. Black areas are regions where the emission is brighter 
than the average and white areas are regions where the emission 
is fainter than the average. In this set of images the rotational 
character of the eastern lobe becomes more clear as the black and white 
flip from December 2011 to February 2013. The color bar shown in the lower right panel is in 
units of MJy~sr$^{-1}$.}
        \label{fig:ObservationGridSubAvg}
\end{figure*}

To more clearly illustrate the variation in the scattered light nebula, we show the 
observations of the protostellar system with the average of all the observed images subtracted 
in Fig. \ref{fig:ObservationGridSubAvg}. We find that  a pattern of variable illumination
is strongly present in the eastern lobe and the areas of greater than average illumination
(and less than average) appear to rotate across the scattered light nebulae from south to north 
with increasing time.
This is in the same direction as the disk and envelope rotation observed by 
\citet{ohashi1997} and \citet{tobin2012}.  The rotating character of the illumination is
not evident in the western lobe, probably due to its overall low intensity. During 2013 and 2014,
the overall intensity of the eastern lobe was lower, causing an overall deficit of emission
when the average is subtracted, but we can still see areas of enhanced and depressed illumination
rotating across the eastern lobe.

\section{Radiative transfer modeling}

The variable illumination and its apparent rotational nature
suggest a connection with the illumination source, which is 
primarily the inner disk and protostar. Thus, the origin of the variability
may be related to the inner disk structure and we conducted radiative transfer modeling
to examine the possible effects of the inner disk that may present themselves
as variable illumination in the scattered light nebula. We used the 
radiative transfer code of \citet{whitney2013}, for which previous versions of the
code \citep{whitney20031} have been used to model L1527 IRS \citep{tobin2008,tobin2010,tobin2013}. The code produces images in 
the near-infrared by propagating photons from the central star, accretion hotspots, and disk that 
can either be absorbed or scattered on their way from the protostar 
and disk out into the envelope. While the code can also 
produce spectral energy distributions, our focus here is on
visualizing the effects of the disk on the overall scattered light images. 

The code now includes the ability
to add perturbations to the disk in the form of gaps, spirals, warps, and/or
a puffed-up inner rim. Furthermore, a misaligned inner disk, with respect to 
the outer disk, is also possible. Since most of the emission in the scattered light nebula
emanates from the inner, optically-thick rim of the dusty disk, we
used a warped inner disk to compare with the observations. 
The remaining parameters are kept unchanged from \citet{tobin2013} as we modify the inner disk
only. From there, we produced images at different azimuthal viewing angles in 15\degr\ increments
to observe how the illumination of the scattered light nebula changed with the rotation of the warped inner disk.

\begin{figure*}
        \centering
        \includegraphics[width=1.0\linewidth]{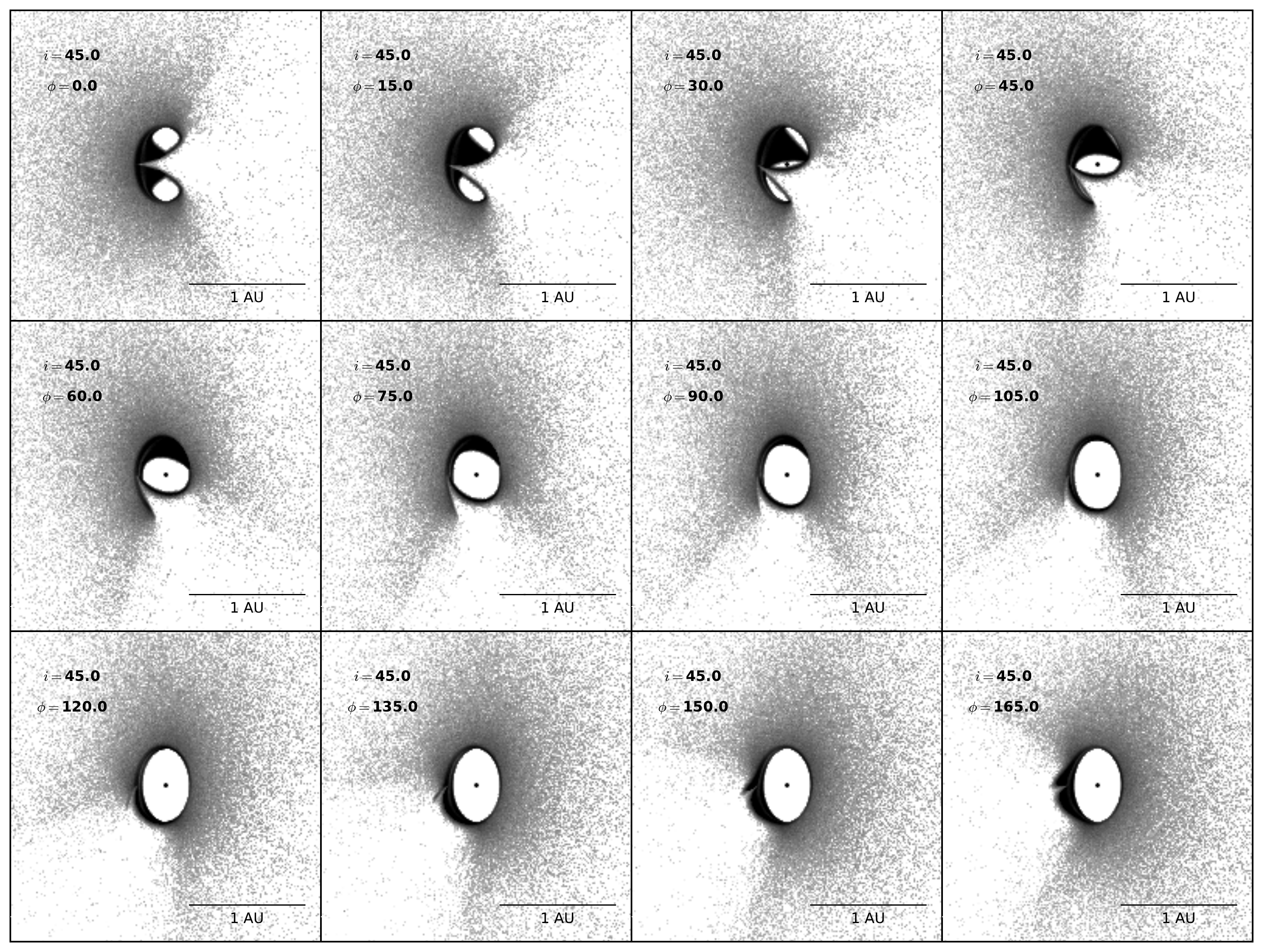}
    \caption{View of the warped inner disk from the radiative transfer model at Ks-band
from an inclination angle of 45\degr\ better shows the structure.
The image is plotted as an inverse grayscale where darker means more intensity, thus the white triangular
region that rotates between images corresponds to shadowing by the inner disk warp.}
        \label{fig:Brian-warp-small-scale-Ks-images-grid}
\end{figure*}

The structure 
of the inner disk warp is visualized in Fig. \ref{fig:Brian-warp-small-scale-Ks-images-grid} where
we computed a model without an envelope, focusing on very small radii to resolve the inner disk.
The warped structure is rather narrow and extends very high vertically from
the disk plane. The warp manifests as a variable scale height $h(\varphi)$ dependent on azimuthal angle such that $h(\varphi) = h_{max}\cos^{n}\varphi$, where $h_{max}$ is the maximum height (see Table \ref{table:warp-largescale}, defined to occur at disk azimuthal angle $\varphi = 0$) and $n$ is the warp exponent. We note that our assumed warp parameters are probably not physically realistic, 
but the exaggerated warp is meant to demonstrate the feasibility
of a warp to produce the variable illumination. 
We found that warps with a large exponent 
(odd integers between 39 and 47, see Table \ref{table:warp-largescale}) were 
the ones that best provided the type of illumination pattern within the outflow 
cavities that could be consistent with the observations. We note, however, that we did not fully explore the 
parameter space, so the parameters of the warp we use in this paper are by no
means a fit to the data. They simply demonstrate the same physical behavior as observed
toward L1527 IRS
and it is likely that other warp parameters could produce similar
results with sufficient exploration of parameter space.

\begin{figure*}
        \centering
        \includegraphics[width=1.0\linewidth]{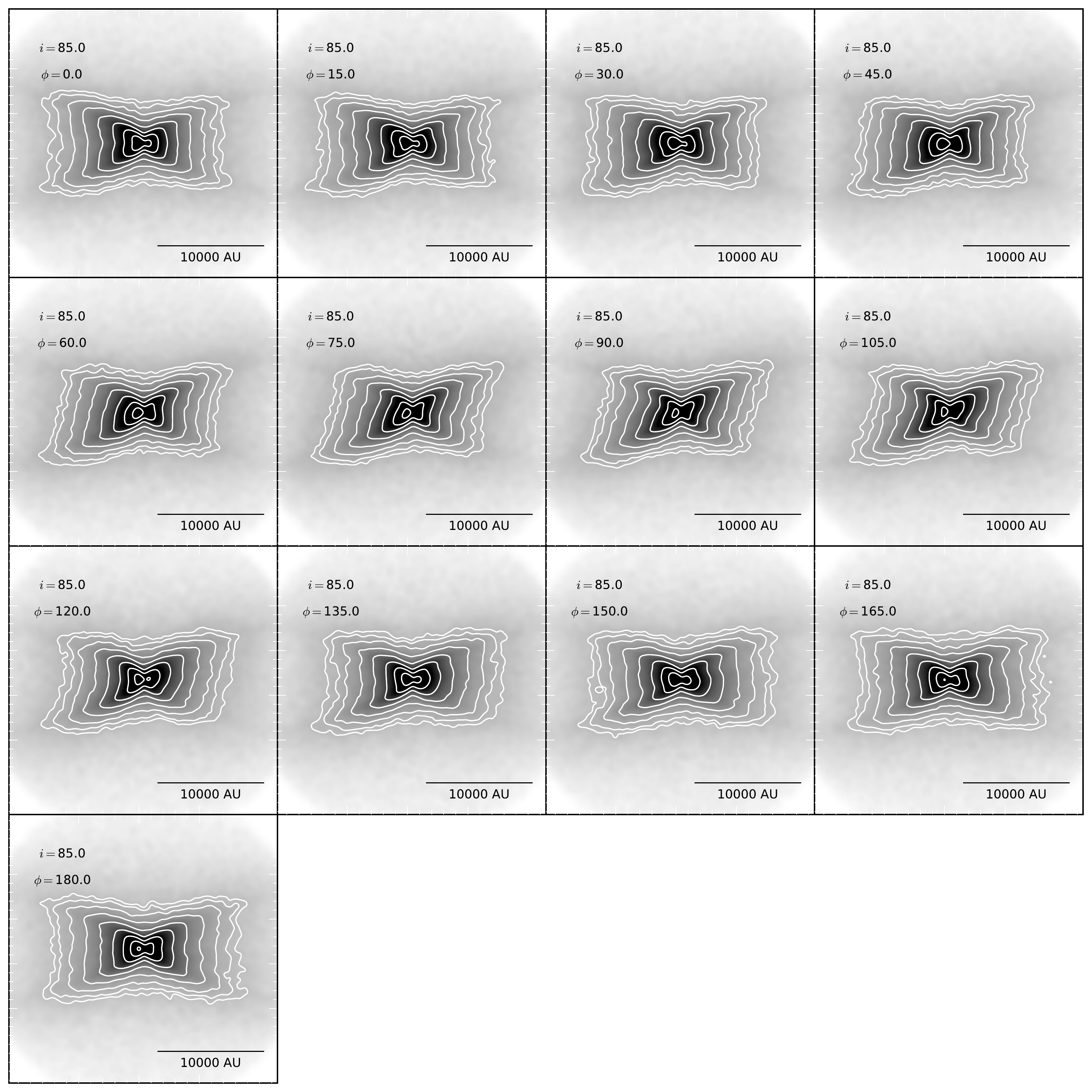}
    \caption{Disk warp simulations viewed on the scale of the protostellar envelope, similar to the
    field of view shown for L1527 IRS in Fig. \ref{fig:ObservationGrid}. These images viewed in 15\degr\ increments of
    azimuthal angle show the effect of a rotating inner disk warp on the scattered light nebula on larger scales. 
    The contour levels are defined as $n = 2^{m}$, where $m$ is the set of integers from -2 to 6 and the units are MJy~sr$^{-1}$.}
        \label{fig:Brian-warp-largescale-Ks-images-grid}
\end{figure*}

Figure \ref{fig:Brian-warp-largescale-Ks-images-grid} shows the effect 
of this warp on the simulated scattered light nebula at larger scales, 
approximately the same scale from which L1527 IRS is viewed in the FanCam observations, as shown in Fig. \ref{fig:ObservationGrid}. 
We view the system in increments of 15\degr\ in the $\phi$ direction
(azimuthal coordinate) to mimic the effect of the inner disk rotating.

\begin{figure*}
        \centering
        \includegraphics[width=1.0\linewidth]{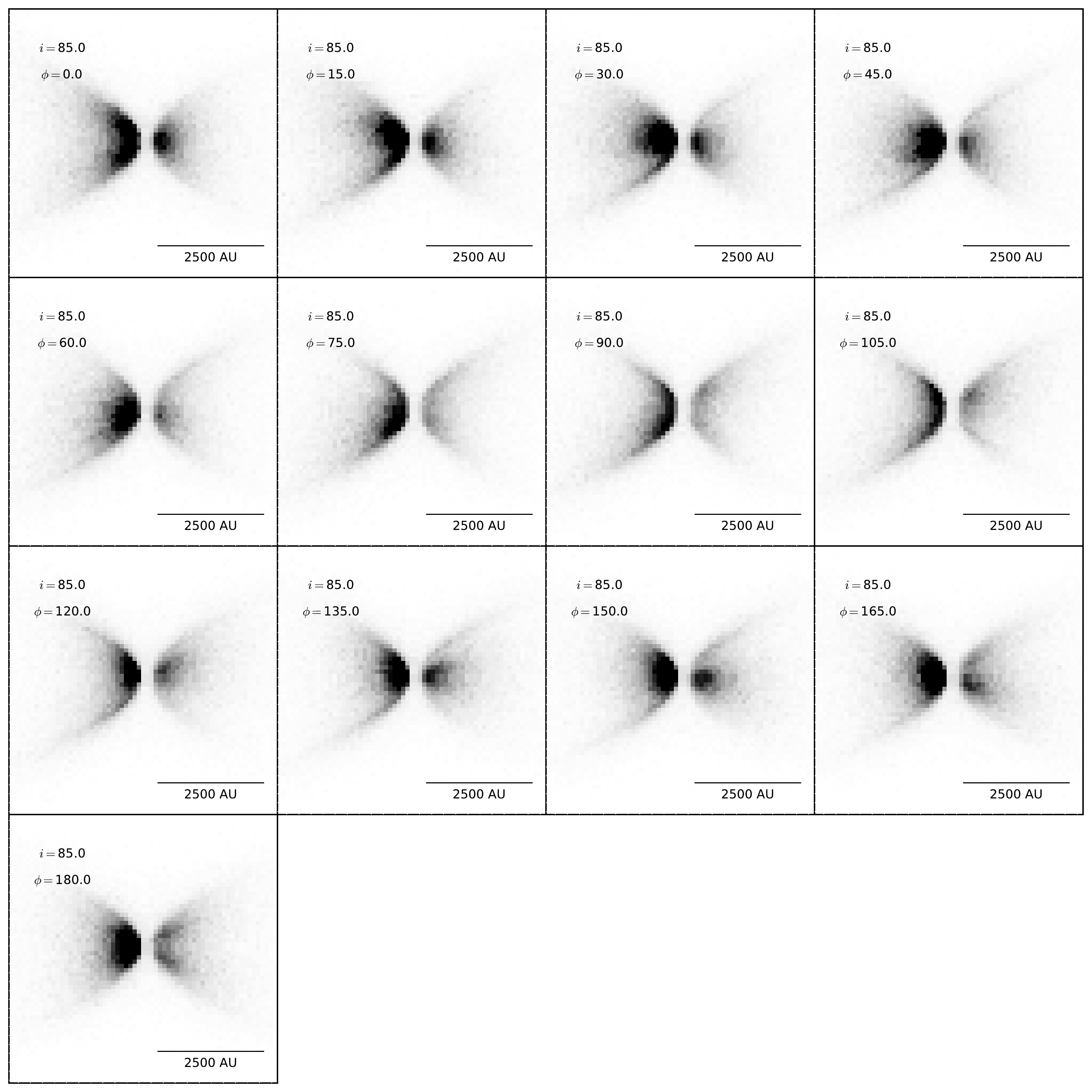}
    \caption{Disk warp simulations viewed on a zoomed-in scale compared to Fig. \ref{fig:Brian-warp-largescale-Ks-images-grid} without smoothing. These images viewed in 15\degr\ increments of
    azimuthal angle show the effect of a rotating inner disk warp on the scattered light nebula on smaller scales.}
        \label{fig:Brian-warp-largescale-Ks-images-grid-zoom}
\end{figure*}

\begin{figure*}
        \centering
        \includegraphics[width=1.0\linewidth]{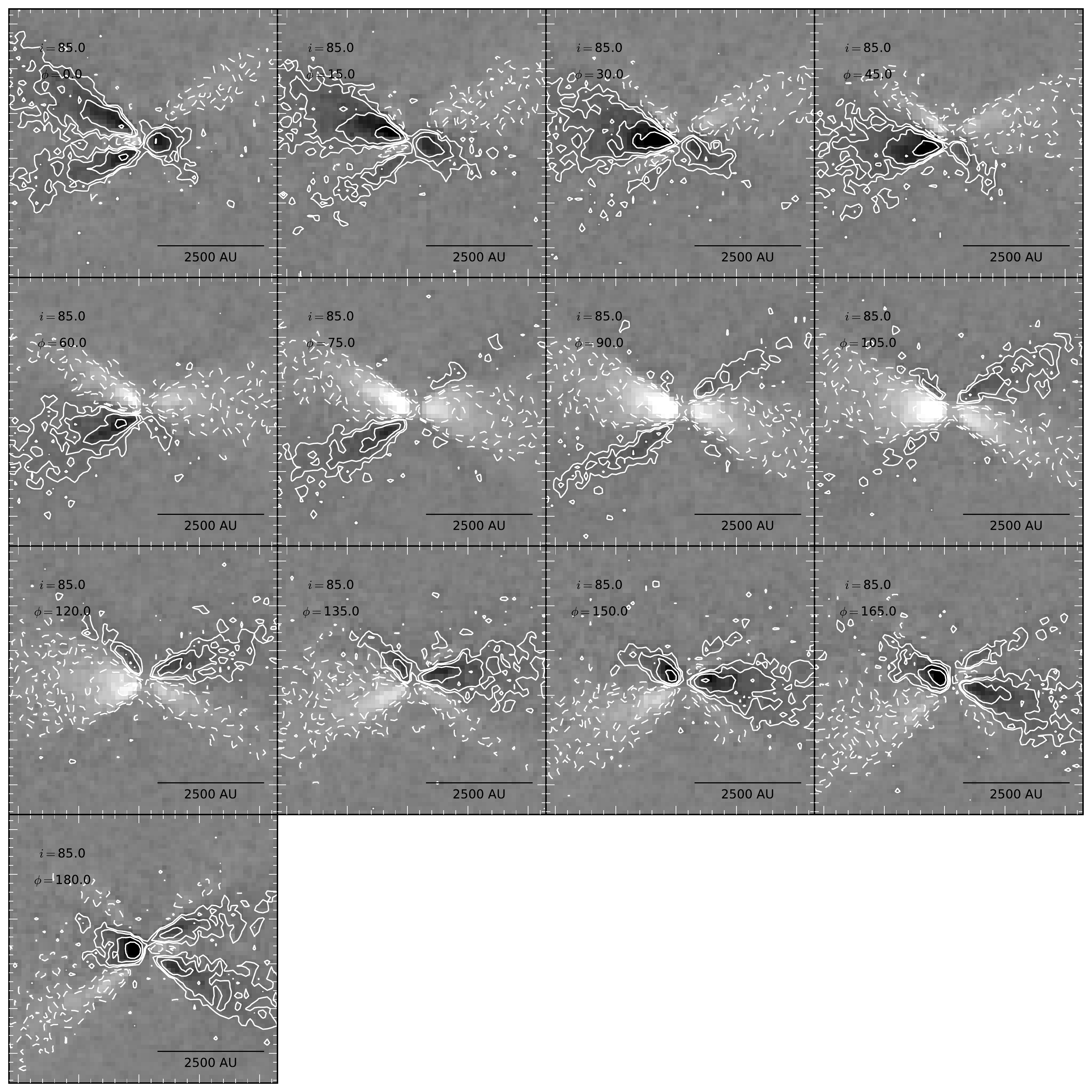}
    \caption{Disk warp simulations viewed on a zoomed-in scale compared to Fig. \ref{fig:Brian-warp-largescale-Ks-images-grid} with the average subtracted out. These images viewed in 15\degr\ increments of
    azimuthal angle show the effect of a rotating inner disk warp on the scattered light nebula on smaller scales. The contour levels are $\pm 2^{m}$, where $m$ is the set of integers from 0 to 3 and the units are MJy~sr$^{-1}$.}
        \label{fig:Brian-warp-largescale-Ks-images-grid-subavg}
\end{figure*}

At $\phi = 0$ to 45\degr\ of Fig. \ref{fig:Brian-warp-largescale-Ks-images-grid} (top row) 
the scattered light morphology is approximately symmetric and there is no
change in the illumination; the same is true for $\phi = 135$ to 180\degr\ (bottom row).
However, for $\phi = 60$ to 120\degr\ (middle row of Fig. \ref{fig:Brian-warp-largescale-Ks-images-grid}),
there is a clear slant in the scattered light images from the southeast to northwest. Furthermore, we did not simulate
($\phi = 195$ to 360\degr) since the illumination pattern would be mirrored. Thus, 
on average this illumination pattern would show a rotating pattern of
increased brightness, which is reflected in Figs. \ref{fig:Brian-warp-largescale-Ks-images-grid-zoom} and \ref{fig:Brian-warp-largescale-Ks-images-grid-subavg}, which show the same simulation outputs at a smaller scale without smoothing or with the average subtracted out, respectively. The latter shows that there is a correlation between intensity and the shearing-type action on larger scales. We do not attempt to fully reproduce the asymmetric intensity of the
eastern and western lobes or the overall decrease of intensity. However, 
this result demonstrates that a warp in the inner disk 
can produce a rotating pattern of illumination variability.

\section{Discussion}

The clear morphological variability of the outflow cavities in the observations
and the apparent rotational nature of the variability strongly suggests
that the origin of the variable illumination is from the disk surrounding the protostar.
The disk has been observed at multiple millimeter and sub-millimeter wavelengths
\citep{maury2010,tobin2012,tobin2013, ohashi2014, aso2017, vanthoff2018}, finding clear evidence
of a dusty, rotationally supported disk. However, the edge-on nature of the system, 
and high opacity of the midplane prevent us from studying the nature of the inner disk
with millimeter interferometry. The likely sub-AU scales of the dust destruction
radius, where most of the emission of the scattered light cavities in L1527 IRS originates,
is still beyond the reach of observatories like the Atacama Large Millimeter/submillimeter Array (ALMA) and the Very Large Array (VLA), even in the most
nearby systems \citep{andrews2016}. Thus, observations of protostellar variability 
provide some of the only hints of the inner disk structure. Many protostars
are shown to have variability in their near to mid-infrared photometry from
the Young Stellar Object VARiability (YSOVAR) survey \citep{stauffer2010, rebull2015}. Several Class 0 systems were found to exhibit modest variability by the YSOVAR survey, and at first glance the variable character of L1527 IRS
may appear different due to the variability being resolved in the scattered light cavities out to $\sim$0.1~pc scales. However, YSOVAR was observing more distant star-forming regions, so it is possible that variability in the scattered light cavities was occurring, but
could not be resolved by \textit{Spitzer}.

\subsection{Possibility of the variability originating from the inner disk}
The variability in L1527 is consistent with having an inner disk origin as
demonstrated with a disk warp model showing a variable illumination pattern. The central protostellar
mass of L1527 is known to be 0.2 to 0.45 $M_{\odot}$ from observations of molecular line kinematics
from the disk \citep{tobin2012,ohashi2014,aso2017, sakai2014}. Our sparse time sampling, with 
significant gaps, makes it difficult to ascertain if (or when) we have observed a complete 
rotation period of the variable illumination. We can see that there was a bright limb on the 
southeast portion of the nebula when we started monitoring in October 2011 
and then this feature is again present at the same location in January 2013. If we assume that this
motion accounts for approximately one orbital period, the length of time is $\sim$15 months or more simply
500 days. If the scattered light variability is related to the rotation period of the inner disk, Kepler's
third law indicates that an orbital period of $\sim$480 days around a 0.35 $M_{\odot}$
protostar corresponds to $\sim$0.85 AU. This points to an origin of variable illumination in the 
inner disk at a distance of $\sim$0.85~AU from the protostar, which is roughly comparable to the radius of the inner disk (see Table \ref{table:warp-largescale}) from which the proposed warp originates. The dust destruction radius,
where the optically thick inner disk begins, is typically defined to be at a temperature of 
$\sim$1400~K. For L1527, the inner radius of the dusty disk is expected to be $\sim$0.15 AU if
the dusty disk extends all the way to the dust destruction radius. However, our estimate of the orbital period has significant uncertainty in that it could be shorter (or longer) than we are 
suggesting here. However, in any event the timescale for the morphological variability
at least appears consistent with the orbital timescales expected to be found
in the inner disk.

\subsection{Other possible mechanisms for variability}

While we did not explore all possibilities to produce the observed variability within the radiative transfer models,
there are other effects beyond a disk warp that could also give rise to variability. 
One possibility that can also produce an asymmetric
brightness distribution in the outflow cavities is a misaligned inner disk.
\citet{matsumoto2017} 
suggested that the misalignment of a protostellar disk could arise 
from the acceleration of the outflows due to a magnetic field. If the protostar within L1527 IRS is 
rotating rapidly, then the star itself will be somewhat oblate, causing the spin axis of the 
star to trail that of the disk, which could cause a disk misalignment \citep{spalding2014}. 
Disk tilting would also cause stochastic wandering of the bipolar outflows. Depending
on the extent of the misalignment, a portion of the inner disk beyond the misaligned portion
would be shadowed and a portion would be more highly illuminated. \citet{wood2001}
showed that misaligned disks in a close binary system could preferentially illuminate one
side of the outflow cavity, but in this case the misaligned inner disk of an assumed
single system could produce the same effect and the brighter portion of the inner disk
beyond the misaligned disk could give rise to the rotating pattern of the scattered light
variations that might have a period of $\sim$1.3~yr. However, this may require a somewhat
fine-tuned system.

Several other mechanisms can result in a misaligned inner disk, most of which are related to
planet-disk interactions, or possibly companion star-disk interactions. Thus, misaligned inner disks
are observed toward gapped protoplanetary disks \citep[e.g., AA Tau; ][]{loomis2017} and others show
shadowing in high-resolution observations of disks that can be attributed to an inner disk \citep{zhu2018}.
It is unclear whether L1527 IRS has a gapped disk (due to a planet or unseen companion) and how far that gap could
extend and still produce a significant amount of scattered light. A misaligned inner disk
alone could not produce an asymmetric illumination pattern because it would require an extremely
large misalignment angle, and the precession of the misaligned inner disk would have to
have a timescale on the order of the $\sim$1.3~yr possible period of the rotating illumination 
pattern. Misaligned inner disks are shown to precess on greater than 100~yr timescales from numerical
simulations \citep{facchini2017}.

It remains possible that there is an unseen binary companion passing above and 
below the disk plane on a short period orbit. This companion could produce an asymmetric
illumination pattern in the 
outflow cavities as well as partially explain the periodic behavior seen 
in the observations. Furthermore, an unseen binary companion could also lead to the possible production
of a misaligned inner disk and/or a warp in the disk. Thus, a combination of multiple effects could
be combined to produce the overall morphology of the outflow cavity illumination and an unseen
companion could naturally produce many of these effects, as has been shown in numerical models \citep[e.g.,][]{zhu2018}.
However, it is unclear if such a close companion could be observationally verified in the
near future. The disk midplane is completely obscured in the near to mid-infrared \citep{tobin2010}, 
and the dust continuum emission is optically thick in the submillimeter and millimeter \citep{vanthoff2018}, effectively
hiding any potential companion. Moreover, VLA observations at 7~mm 
are the highest resolution possible at present for more optically thin dust emission and these data only show a disk and do not have the resolution 
to examine structures on $\sim$1AU scales (Melis et al. in prep.). The enhanced Multi Element Remotely Linked Interferometer Network (e-MERLIN) could be used for higher resolution, but only if L1527 IRS is bright (and compact) enough at those resolutions.

\subsection{Additional examples of scattered light variability}

System L1527 IRS is just one example of what may be a class of objects showing resolved
time variability in their scattered light nebulae. One of the original variable scattered
light nebulae from R Mon \citep[also known as Hubble's Variable Nebula,][]{hubble1917} is also
found around a young, but massive star \citep{fuente2006}. There has also been periodic variability observed
in the protostar Per-emb-28 (LRLL 54361), which was attributed to a binary interaction
resulting in pulsed accretion on an $\sim$25 day period \citep{muzerolle2013}. However, there are other examples toward
more similar low-mass protostars that are more similar to the slow
changes observed in L1527 IRS.

System L483 IRS as reported by
\citet{connelley2009} is the most directly comparable system to L1527 IRS. It
is also a Class 0/I protostar viewed nearly edge-on and both
cavities are visible. L483 IRS has similar variability characteristics to L1527 IRS as well. It has 
resolved morphological variability, variability in overall amplitude, and 
independent variations in the bipolar scattered light cavities. However, L483 IRS has
not been subject to the more frequent monitoring that we conducted for L1527 IRS; the observations of L483
are spread over many months to years apart. \citet{connelley2009} suggested opaque clouds
within $\sim$1~AU of the protostar could be the cause of the variability, but such small-scale features could also be associated with an inner disk. 

The behavior of L1527 IRS is also analogous to the disk around HH30, but HH30 shows clear variations
in scattered light disk rather than the large-scale nebula.
\citet{watson2007} discuss many possible mechanisms to produce the variation, including
variation that only affects one side of the nebula. They conclude that a combination of
overall variability and disk asymmetries (including a possible warp) could be responsible
for the variability in the scattered light disk.

We also mentioned the YSOVAR studies that find numerous variable protostellar and
pre-main sequence stars with disks that show variability with a variety of light curves \citep{stauffer2010}.
These studies could be revealing the types of inner disk variability that might give
rise to the morphological variability in the scattered light nebula toward L1527 IRS and other protostars.
L1527 IRS and L483 IRS are not likely to be the only examples of variable scattered light nebulae, and the 
data already exist to enable further discovery of these types of objects using archival \textit{Spitzer}
and \textit{Wide-field Infrared Survey Explorer} data.

\section{Conclusions}

We have presented ground-based near-infrared monitoring of the Class 0 protostar
L1527 IRS finding variability within its scattered light nebula
through observations at the Ks-band taken over the course of seven years, and higher
cadence monitoring that was conducted over three years. We find variability in the overall surface
brightness of the scattered light nebula from the protostar, independent
variability in both of the bipolar outflow cavity lobes, and variability in the illumination pattern
of the cavities that appears to rotate
across the eastern outflow cavity. We have demonstrated that an inner disk
warp can produce a variable pattern that would rotate across
the outflow cavity as the inner disk orbits the protostar.

Other mechanisms such as a misaligned inner disk and/or an unseen binary
companion could also possibly produce the observed variable illumination in the 
scattered light nebulae, but we aimed to produce the effect with a
minimal set of parameters. Higher and more uniform cadence monitoring
with a larger telescope with better seeing could lead to an 
improved characterization of the variability in L1527. However, regardless of
the details, we conclude that the origin of the variable illumination pattern
is tied to the inner disk, and time domain monitoring of protostellar and
protoplanetary systems provides important and unique constraints on the
structure of the inner disks around protostars and pre-main sequence stars.

\begin{acknowledgements}
We would like to thank the anonymous referee for their helpful comments that improved the quality of the manuscript. BTC and JJT wish to thank L. Hartmann and N. Calvet for suggesting
the collaboration while BTC was studying at Leiden University. We acknowledge
the support from the University of Michigan to the MDM observatory and the University
of Virginia in their support of the Fan Mountain observatory.
We thank H. Borsich for helping JJT get started observing on FanCam, as well as
A. Mead and J. Heartly for helping with the observations.
The National Radio Astronomy Observatory is a facility of the National 
Science Foundation operated under cooperative agreement by Associated Universities, Inc.
\end{acknowledgements}

\clearpage

\onecolumn
\begin{longtable}{llllll}
\caption{\label{table:observation-log} L1527 observation log.}\\
\hline\hline
Date & Telescope/Instrument & Integration time & Airmass & Seeing & Standard deviation \\
(UT) & & (m) & ($\sec(z)$) & ($\arcsec$) & (MJy~sr$^{-1}$) \\
\hline
\endfirsthead
2007 December 28 & Hiltner 2.4m/TIFKAM           & 35                         & 1.03            & 1.0   & 0.0561 \\
2009 December 03 & Hiltner 2.4m/TIFKAM           & 30                         & 1.06 - 1.16            & 0.9   &  0.0446 \\
2011 October 07 & 31~in/FANCAM                 & 65                         & 1.03 - 1.1            & 1.3    & 0.0298 \\
2011 October 23 & 31~in/FANCAM                 & 95                         & 1.24 - 1.04            & 1.2   & 0.0262  \\
2011 November 07 & 31~in/FANCAM                 & 100                         & 1.3 - 1.05            & 1.2  &  0.0276  \\
2011 December 11 & 31~in/FANCAM                 & 100                         & 1.2 - 1.05            & 2.0  &  0.256  \\
2012 January 07 & 31~in/FANCAM                 & 115                         & 1.28 - 1.03            & 2.25  &  0.0255  \\
2012 January 31 & 31~in/FANCAM                 & 100                         & 1.09 - 1.04            & 1.75  &  0.0307  \\
2012 February 18 & 31~in/FANCAM                 & 105                         & 1.02 - 1.15            & 1.8  &  0.0258  \\
2012 March 02 & 31~in/FANCAM                 & 70                         & 1.16 - 1.46            & 1.75    & 0.0364 \\
2012 March 15 & 31~in/FANCAM                 & 140                         &  1.1 - 2.2            & 1.5     & 0.0328 \\
2013 January 21 & 31~in/FANCAM                 & 110                         & 1.11 - 1.03            & 1.75  &  0.0455  \\
2013 February 21 & 31~in/FANCAM                 & 110                         & 1.02 - 1.18           & 2.5  &  0.0310  \\
2013 April 14 & 31~in/FANCAM                 & 50                         & 1.65 - 2.4            & 2.0      & 0.0343 \\
2013 April 21 & 31~in/FANCAM                 & 50                        & 1.95-3.3            & 2.5      & 0.0334 \\
2013 October 02 & 31~in/FANCAM                 & 95                         & 1.17 1.02            & 1.25    & 0.0250 \\
2013 November 04 & 31~in/FANCAM                 & 105                         & 1.24 - 1.03            & 1.8  &  0.0242  \\
2013 December 19 & 31~in/FANCAM                 & 105                         & 1.22 - 1.03            & 1.4  &  0.0224  \\
2014 January 20 & 31~in/FANCAM                 & 130                        & 1.15 - 1.02            & 2.8   & 0.0230  \\
2014 February 28 & 31~in/FANCAM                 & 105                         & 1.02 - 1.2            & 2.0  &  0.0285  \\
\hline
\end{longtable}

\begin{longtable}{lll}
\caption{\label{table:warp-largescale} Warped disk model parameters, most of which have been unchanged from the best fit model in \cite{tobin2013}. Parameters most relevant to this result are in boldface and parameters that have been adjusted from the value in the best fit model are italicized. The stellar mass
is only used in setting the density structure, changing this to the measured protostar mass would
require increasing the envelope infall rate to maintain the same density structure.}\\
\hline\hline
Parameters & Description & Values \\
\hline
\endfirsthead
        $\gamma$'s & \textbf{Number of photons} & $1\times10^{8}$ \\
        $R_{\star}(R_{\odot})$ & Stellar radius & 2.09 \\
        $T_{\star}(K)$  & Stellar temperature & 4000 \\
        $M_{\star}(M_{\odot})$ & Stellar mass & 0.5 \\
        $M_{disk}(M_{\odot})$ & Disk mass & 0.005 \\
        $h$ (100 AU) & Disk scale height at 100 AU & 48.0 \\
        $H_{0}$ ($R_{\star}$) & Disk scale height at $R_{\star}$ & 0.03 \\
        $\alpha$ & Disk radial density exponent & 2.5 \\
        $\beta$ & Disk scale height exponent & 1.3 \\
        $\dot{M}_{disk}$ ($M_{\odot}$ yr$^{-1}$) & Disk accretion rate & $1.5\times 10^{-6}$ \\
        $R_{trunc}$ ($R_{\star}$) & Magnetosphere co-rotation radius & 3.0 \\
        $F_{spot}$ & Fractional area of accretion hotspot & 0.01\\
        $R_{disk, min}$ ($R_{\star}$, AU) & Disk inner radius & 14.25, 0.14\\
        $R_{disk, max}$ (AU) & Disk outer radius & 125\\
        $R_{env, min}$ ($R_{\star}$)  & Envelope inner radius & 42.75 \\
        $R_{env, max}$ (AU) & Envelope outer radius & 15000\\
        $\dot{M}_{env}$ ($M_{\odot}$ yr$^{-1}$) & Envelope mass infall rate & $4.5\times10^{-6}$\\
        $\rho_{1 AU}$ (g cm$^{-3}$) & Envelope density at 1 AU & $7.25\times10^{-14}$\\
        $b_{out}$ & Outer cavity shape exponent & 1.5\\
        $\theta_{open, out}$ (degrees) &Outer cavity opening angle & 20\\
        $\rho_{c}$ (g cm$^{-3}$) &Cavity density& 0\\
        $\beta_{dust, mm}$ & Millimeter dust spectral index & 0.25 \\
        Warp Height & \textbf{\textit{Multiplicative factor of warp scale height above disk scale height}} & 10.0 \\
    $n$ & \textbf{\textit{Exponent for azimuthal disk warp ($\cos^{n}$)}} & 41 \\
        $\theta$ (degrees) & Inclination angles of high S/N images & 85.0 \\
        $\phi$ (degrees) & Azimuthal angles & 0,15.0,30.0,\dots,180.0\\
        $\lambda$ ($\mu$m) & wavelength & 2.15  \\
\hline
\end{longtable}

\clearpage

\bibliographystyle{aa}
\bibliography{Draft}

\end{document}